\documentclass[a4paper,11pt]{article}
\usepackage{pos}
\usepackage{graphicx,psfrag,epsf,color}
\usepackage{amsmath,amsfonts}
\usepackage{physics}
\usepackage{siunitx}
\usepackage{subfig}

\makeatletter
\g@addto@macro\normalsize{%
  \setlength\abovedisplayskip{5pt}
  \setlength\belowdisplayskip{5pt}
  \setlength\abovedisplayshortskip{5pt}
  \setlength\belowdisplayshortskip{5pt}
}

\usepackage{titlesec}
\titlespacing*{\section}{0pt}{0.3\baselineskip}{0.3\baselineskip}
\titlespacing*{\subsection}{0pt}{0.3\baselineskip}{0.3\baselineskip}

\title{New Probes of Electron--Muon Universality in $\boldmath{B \to K\ell^+\ell^-}$ Decays}
\author[a,b]{Robert Fleischer}
\author[a,c]{Eleftheria Malami}
\author*[a,b]{Anders Rehult}
\author[a,d]{K. Keri Vos}

\affiliation[a]{Nikhef, Science Park 105, NL-1098 XG Amsterdam,  Netherlands}

\affiliation[b]{Department of Physics and Astronomy, Vrije Universiteit Amsterdam, NL-1081 HV Amsterdam, Netherlands}

\affiliation[c]{Center for Particle Physics Siegen (CPPS), Theoretische Physik 1, Universität Siegen, D-57068 Siegen, Germany}

\affiliation[d]{Gravitational Waves and Fundamental Physics (GWFP), Maastricht University, Duboisdomein 30, NL-6229 GT Maastricht, the Netherlands}

\emailAdd{arehult@nikhef.nl}

\abstract{In the pursuit of physics beyond the Standard Model, a promising path is the study of B-meson decays caused by the transition $b \to s\ell^+\ell^-$. A key observable in such decays is the ratio $R_K$, which measures electron--muon universality in $B \to K \mu^+\mu^-/e^+e^-$. At first sight, the recent LHCb measurement of $R_K \sim 1$ may seem to largely constrain deviations from universality in these decays. However, we show that this is actually not the case: new sources of CP violation allow for significant universality violation consistent with $R_K \sim 1$. This provides an exciting new opportunity to search for New Physics by measuring differences between CP asymmetries in $B \to K\mu^+\mu^-$ and $B \to K e^+e^-$.}

\FullConference{20th International Conference on B-Physics at Frontier Machines (Beauty2023)\\
 3-7 July, 2023\\
Clermont-Ferrand, France\\}

\begin{document}
\maketitle

\section{Introduction}
Do New Physics (NP) effects beyond the Standard Model (SM) discriminate between different lepton flavours? This question has been explored through the measurement of lepton--flavour universality ratios, most prominently through \cite{Hiller:2003js}
\begin{equation}
    R_K  \equiv \frac{ \Gamma(B^-\to K^-\mu^+\mu^-) +\Gamma(B^+\to K^+\mu^+\mu^-) }{\Gamma(B^-\to K^-e^+e^-)  + \Gamma(B^+\to K^+e^+e^-) } \ .
    \label{eq:RKav}
\end{equation}
For several years measurements of this observable deviated from the SM value of $1$ \cite{LHCb:2014vgu,LHCb:2019hip,LHCb:2021trn,Bordone:2016gaq}. Recently, however, $R_K$ was measured by the LHCb collaboration to be consistent with the SM within one standard deviation \cite{LHCb:2022qnv,LHCb:2022zom}. At first sight, this new result seems to strongly limit violations of electron--muon universality in $B \to K\ell^+\ell^-$ decays. However, puzzling tensions still exist in data on $B^+ \to K^+\mu^+\mu^-$ and other, related $b \to s\mu^+\mu^-$ decays (see e.g. \cite{Capdevila:2023yhq} for a recent review). Given this situation, is there still space left for electron--muon universality violation?

\section{Charting the parameter space for electron--muon universality violation}
The low-energy effective Hamiltonian for $b \to s \ell^+ \ell^-$ decays is
\begin{equation}\label{eq:ham}
    \mathcal{H}_{\rm eff} = - \frac{4 G_F}{\sqrt{2}} \left[\lambda_u \Big\{C_1 (\mathcal{O}_1^c - \mathcal{O}_1^u) + C_2 (\mathcal{O}_2^c - \mathcal{O}_2^u)\Big\} + \lambda_t \sum\limits_{i \in I} C_i \mathcal{O}_i \right] \ ,
\end{equation}
where $\lambda_q = V_{qb} V_{qs}^*$ and $I = \{1c, 2c, 3, 4, 5, 6, 8, 7^{(\prime)}, 9^{(\prime)}\ell, 10^{(\prime)}\ell, S^{(\prime)}\ell, P^{(\prime)}\ell, T^{(\prime)}\ell\}$.
For simplicity we consider only NP in the coefficient $C_{9\ell} \; (\ell = \mu,e)$, whose operator is defined as
\begin{equation}
    \mathcal{O}_{9\ell} = \frac{e^2}{(4\pi)^2} [\bar s \gamma^\mu P_{L} b] (\bar \ell \gamma_\mu \ell) \ .
\end{equation}
For a broader discussion including $C_{10\ell}$ see \cite{Fleischer:2023zeo}, where we also discuss our treatment of the relevant form factors and hadronic long-distance effects. 

We first constrain the muonic coefficient $C_{9\mu}$ using experimental data. We then use the new $R_K$ measurement to study by how much $C_{9e}$ can differ from $C_{9\mu}$. We perform this procedure twice, first assuming the Wilson coefficients to be real numbers and then allowing them to be complex, thereby opening the door to new sources of CP violation.

\subsection{Real Wilson coefficients}
\begin{figure}
    \centering
    \includegraphics[width=0.28\textwidth]{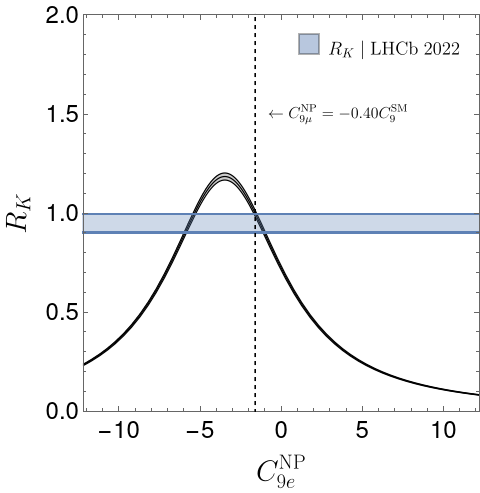}
    \caption{The ratio $R_K$ as a function of a real $C_{9e}^{\rm NP}$, corresponding to no new sources of CP violation.}
    \label{fig:fig1}
\end{figure}

To constrain a real $C_{9\mu}$, we use the most recent data on the branching ratio $\mathcal{B}(B^+ \to K^+\mu^+\mu^-)$ \cite{LHCb:2014cxe}.  To accommodate these data within $1 \sigma$, we find that $C_{9\mu}$ needs to take a value within
\begin{equation}
\begin{aligned}\label{eq:range}
C_{9\mu}^{\rm NP} = [-1.32, -0.40]C_9^{\rm SM} \ .
\end{aligned}
\end{equation}
Fixing $C_{9\mu}$ within this range, we use the recent $R_K$ measurement to calculate the allowed values for $C_{9e}$. Fig.~\ref{fig:fig1} shows the result. The curve indicates $R_K$ as a function of $C_{9e}$, the horizontal band the recent $R_K$ measurement, and the dashed vertical line the value that $C_{9\mu}$ is fixed to. Within the band $C_{9e}$ is constrained to one of two values: it can either take the same value as $C_{9\mu}$, respecting universality, or assume a different, more negative value. Consequently, with real Wilson coefficients, the recent $R_K$ measurement leaves little space for violations of electron--muon universality.

\subsection{Complex Wilson coefficients}
\begin{figure}
    \centering
    \subfloat{\includegraphics[width=0.27\textwidth]{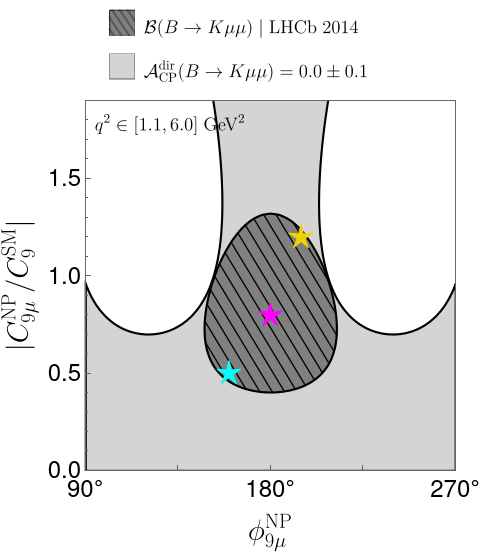} \label{fig:fig2a}}
    \subfloat{\includegraphics[width=0.27\textwidth]{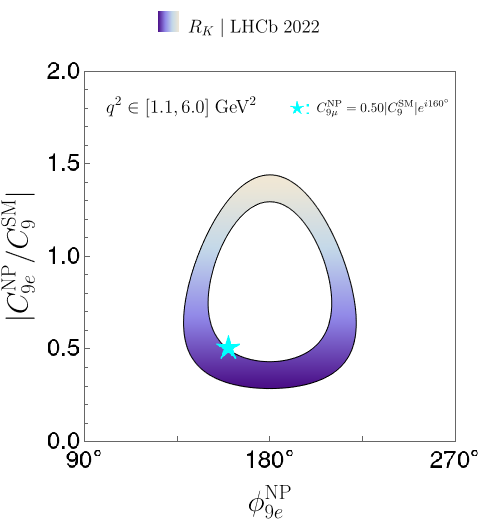} \label{fig:fig2b}}
    \subfloat{\includegraphics[width=0.27\textwidth]{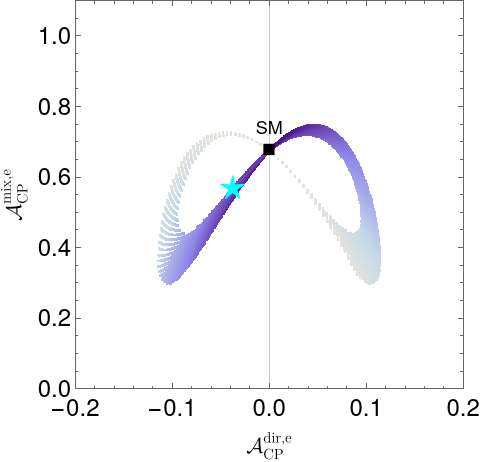} \label{fig:fig2c}}
    \caption{Constraints on a complex $C_{9\mu}^{\rm NP}$ (left), $C_{9e}^{\rm NP}$ (middle), and CP asymmetries in $B \to Ke^+e^-$ (right).}
\label{fig:fig2}
\end{figure}
We constrain a complex $C_{9\mu}$ in Fig.~\ref{fig:fig2a} by using the branching ratio and direct CP asymmetry of $B^+ \to K^+ \mu^+ \mu^-$. Fixing $C_{9\mu}$ to the blue star (see \cite{Fleischer:2022klb} for a determination of $C_{9\mu}$), we use the new $R_K$ measurement to constrain $C_{9e}$. The resulting bound is shown in Fig.~\ref{fig:fig2b}. If NP respects electron--muon universality, then $C_{9e}$ will take the same value as $C_{9\mu}$, i.e. the blue star. However, Fig.~\ref{fig:fig2b} shows that this is not necessarily the case. Instead, $C_{9e}$ can take any value within the egg-shaped region, thereby leaving a surprising amount of space for universality violation.

To obtain the full picture, measuring $R_K$ is not sufficient. We also need measurements of CP asymmetries in $B \to K\mu^+\mu^-$ and $B \to K e^+e^-$. Complex, non-universal Wilson coefficients can cause these CP asymmetries to differ significantly from each other. Fig.~\ref{fig:fig2c} shows the parameter space allowed within the bounds of Fig.~\ref{fig:fig2b} for a direct and a mixing-induced CP asymmetry of $B^0_d \to K_S e^+e^-$, a decay related to $B^+ \to K^+ e^+ e^-$ through isospin symmetry. We could access much of this space given data on $\mathcal{A}_{\rm CP}^{\rm dir}(B^0_d \to K_S e^+e^-)$ (or the related $\mathcal{A}_{\rm CP}^{\rm dir}(B^+ \to K^+e^+e^-)$), which would draw a vertical band in the figure. And we could reach the remaining space with data on $\mathcal{A}_{\rm CP}^{\rm mix}(B^0_d \to K_S e^+e^-)$, which would draw a horizontal band. If either band were to exclude a known $C_{9\mu}$ point (blue star), we would have a clear signal of electron--muon universality violation.

\section{Conclusions}
In light of the recent $R_K$ measurement, we have charted the remaining parameter space for violations of electron--muon universality in $B \to K \ell^+\ell^-$ decays. We have found that there remains a significant amount of unexplored space linked to new sources of CP violation. This space can be explored by searching for differences between CP asymmetries in $B \to K\mu^+\mu^-$ and $B \to K e^+e^-$ decays, providing an exciting new opportunity to reveal NP effects in the coming high-precision era.

\clearpage
\acknowledgments{A.R. would like to thank the organizers for the invitation to the enjoyable conference. This research has been supported by the Netherlands Organisation for Scientific Research (NWO). }

\vspace{0.5em}
\bibliographystyle{JHEP} % bst file
\bibliography{refs.bib}

\end{document}